\documentclass[aps,prb,twocolumn,superscriptaddress,showpacs,amsmath,amssymb,longbibliography]{revtex4-1}
\usepackage[english]{babel}
\usepackage{amsmath}
\usepackage{bm}
\usepackage{graphicx,bbm} 
\usepackage{times}
\usepackage{epsfig} 
\usepackage{soul}
\usepackage[utf8]{inputenc}
\usepackage[colorlinks,linkcolor=blue,citecolor=blue,urlcolor=blue]{hyperref}
\usepackage{color}
\usepackage{latexsym}
\usepackage{ulem}
\definecolor{nred}{RGB}{224,0,0}
\definecolor{nblue}{RGB}{28,130,185}
\definecolor{dgreen}{RGB}{78,138,21}
\definecolor{norange}{RGB}{230,120,20}

\begin{document} 
\title{Slow diffusion and Thouless localization criterion in modulated spin chains}
\author{P. Prelov\v{s}ek}
\affiliation{Jo\v{z}ef Stefan Institute, SI-1000 Ljubljana, Slovenia}
\affiliation{Faculty of Mathematics and Physics, University of Ljubljana, SI-1000 Ljubljana, Slovenia}
\author{J. Herbrych}
\affiliation{Institute of Theoretical Physics, Faculty of Fundamental Problems of Technology, Wroc\l{a}w University of Science and Technology, 50-370 Wroc\l{a}w, Poland}
\author{M. Mierzejewski}
\affiliation{Institute of Theoretical Physics, Faculty of Fundamental Problems of Technology, Wroc\l{a}w University of Science and Technology, 50-370 Wroc\l{a}w, Poland}

\date{\today}
\begin{abstract}
In recent years the ergodicity of disordered spin chains has been investigated via extensive numerical studies of the level statistics or the transport properties. However, a clear relationship between these results has yet to be established. We present the relation between the diffusion constant and the energy-level structure, which leads to the Thouless localization criterion. Together with the exponential-like dependence of the diffusion constant on the strength of quasiperiodic or random fields, the Thouless criterion explains the nearly linear drift with the system size of the crossover/transition to the nonergodic regime. Moreover, we show that the Heisenberg spin chain in the presence of the quasiperiodic fields can be well approached via a sequence of simple periodic systems, where diffusion remains finite even at large fields.
\end{abstract}
\maketitle

\section{Introduction}
The many-body localization (MBL) is the phenomenon which should persist in a disordered quantum system \cite{anderson58} even in the presence of many-body (MB) interaction \cite{basko06}. Due to the absence of thermalization and ergodicity, the existence of the MBL regime should fundamentally change the statistical description of such system. In last decade several features of the MBL have been established, mostly in numerical studies of the anisotropic Heisenberg Hamiltonian with random local fields, i.e., the change in level statistics and spectral properties \cite{oganesyan07,luitz15,serbyn16,suntajs20,sierant20}, logarithmic growth of entanglement entropy \cite{znidaric08,bardarson12,serbyn15}, vanishing of DC transport even at high temperature $T$ \cite{berkelbach10,barisic10,agarwal15,barlev15,steinigeweg16,prelovsek17}, and generally nonergodic correlations \cite{pal10,serbyn13,huse14,luitz16,mierzejewski16}. While such markers are quite evident in finite systems at large disorders, there remains a fundamental question whether the MBL remains stable in the thermodynamic limit and long times \cite{suntajs20,suntajs_bonca_20,sels2020,sels21,vidmar21,krajewski22,Sels_2022}, requiring a well defined MBL phase transition in contrast to quite sharp crossover to a glassy MBL phase. Although indications for the latter is already exponential-like dependence of DC transport on the disorder strength \cite{barisic10,barisic16,steinigeweg16,prelovsek17,prelovsek21,herbrych22}, the debate has been recently stimulated also by other results concerning the MBL transition \cite{bera17, panda20,sierant20,sierant2020_1,abanin21}, spectral properties \cite{suntajs20,vidmar21,sierant22} and avalanche instability \cite{sels21,morningstar22}. 

Studies of random spin chains suffer from large sample-to-sample fluctuations of evaluated quantities for numerically available sizes \cite{herbrych22,vidmar21,krajewski22}. On the other hand, there seems to be numerical evidence \cite{iyer13,barlev17,khemani17,setiawan17,bera17,znidaric18,zhang18,agrawal20,aramthottil21,singh21,strkalj21,sierant22} that analogous models with quasiperiodic (QP) field $W$ should exhibit the MBL at large $W$, although this could be of another universality class \cite{khemani17,znidaric18,morningstar22}. Still, in such model fields are deteministic and no sample averaging is required, as it will be verified also in the present work. Furthermore, most experimental evidence for MBL comes from the studies of the cold-atom systems where the QP potentials are realized \cite{schreiber15,luschen17}.

The central message of this work is that one can derive within the framework of random-matrix theory (RMT) \cite{brody81,wilkinson90,dalessio16} a simple relation between the spin diffusion constant ${\cal D}_0$ and the level-sensitivity parameter $R$, which quantifies the shift of energy-levels due to modified boundary conditions. We refer to this relation as the Thouless localization criterion \cite{edwards72}. Most importantly, this relation explains specific dependences between the system size $L$ and the threshold disorder (or the strength of QP potential) $W^*$ when the system's properties start to deviate from RMT. The latter dependence is linear for random systems, $W^* \propto L$, \cite{suntajs20,suntajs_bonca_20} and sublinear for QP chains \cite{aramthottil21}. Furthermore, we can directly relate this observation to our numerical results for ${\cal D}_0$ which reveal an exponential-like dependence on the disorder strength \cite{barisic10,steinigeweg16,barisic16,prelovsek17,prelovsek21,herbrych22} or quasiperiodic field $W$ (this work). This relation puts relevant restrictions to presumed MBL transition. According to arguments by Edwards and Thouless~\cite{edwards72}, the latter would require the decrease of parameter $R$ with $L$ which we do not observe up to the largest $W$ accessible to our numerical methods, although we cannot exclude such possibility at considerably larger $W$.

\section{Model}
We study a spin chain described by the XXZ model with modulated magnetic field,
\begin{equation}
H = \sum_i \left[ \frac{J}{2}( S^+_{i+1} S^-_i + \mathrm{H.c.}) + J \Delta S^z_{i+1} S^z_i + h_i S_i^z \right]\,, \label{hm}
\end{equation}
where $S^{\pm,z}$ are spin $s=1/2$ operators and we take further on \mbox{$J=1$}. In the equivalent chain of spinless fermions, the anisotropy $\Delta$ represents the strength of the two-body interaction. We work with finite systems of length $L$ and periodic boundary conditions (PBC). In order to be compatible with PBC, we restrict ourselves to commensurate $h_i=(W/2) \cos( 2 \pi k \,i + \phi_0)$, where $k= M/L$ with integer $M$.
 
At high temperatures $T \gg 1$ and for strong fields \mbox{$W >2$} the properties of the studied system crucially dependent on the periodicity of $h_i$. It is well known that for QP golden-mean value $\tilde k = (\sqrt{5}-1)/2$ and in the absence of interaction ($\Delta=0$), the model (\ref{hm}) represents the Aubry-Andre chain \cite{aubry80} having all states localized at $W >2$. The latter is the starting point for most MBL studies at $\Delta \ne 0$ \cite{iyer13,naldesi16,barlev17,khemani17,setiawan17,bera17,znidaric18,zhang18,agrawal20,aramthottil21,singh21,strkalj21,sierant22}. We note that majority of the numerical studies of MBL were performed for $\Delta=1$, where (at $L \sim 20$) the characteristic MBL crossover is found at $W^* \simeq 3$ \cite{naldesi16,setiawan17}, but with noticeable finite-size shift \cite{aramthottil21}. It is evident that such incommensurate value $\tilde k$ is quite close to simpler periodic cases with integer periodicity $P=1/k \ll L$. In the following we study an equivalent QP modulation $k = 1- \tilde k = (3-\sqrt{5})/2 \simeq 0.38$, approximated by closest rational values of $k$, as well as by simple periodic $P=2, 3$ systems.

\section{Level statistics and transport properties} 

Since most studies of MBL rely on the level statistics, it is desirable to connect the transport properties with indicators of RMT. To this end, we study $T \to \infty$ dynamical spin diffusivity, ${\cal D}(\omega)$, related to the spin conductivity $\sigma(\omega)$,
\begin{equation}
{\cal D}(\omega) = \frac{\sigma(\omega)}{\chi_0} = \frac{\pi}{ L \tilde \chi_0 N_{st} } \sum_{m\ne n} |j_{mn}|^2 \delta(\omega- \epsilon_m + \epsilon_n). 
\label{dom}
\end{equation}
Here, we introduced the spin current operator $j = (J/2) \sum_l( i S^+_{l+1} S^-_l + \mathrm{H.c.})$ and its matrix elements (ME) $j_{mn} = \langle m| j | n\rangle$ for the MB eigenstates $|n \rangle, |m\rangle $ with corresponding energies $\epsilon_n,\epsilon_m$, respectively. $N_{st}$ in Eq.~(\ref{dom}) is the dimension of the MB Hilbert space and $\chi_0 = \tilde \chi_0/T$ is the spin susceptibilty. We concentrate on unpolarized spin systems with total $S^z_{tot} \simeq 0$ and use the high-$T$ value $\tilde \chi_0 =1/4$. It is important to differentiate between dynamical diffusion ${\cal D}(\omega)$, Eq. (\ref{dom}), and the spectral function studied for $S^{z}_i$ in the context of the ETH \cite{Luitz2016, serbryn17}. The ME of the spin current are very different from ME of $S^{z}_i$. As a consequence, the spectral function for $S^{z}_i$ shows a pronounced maximum at $\omega \to 0$, whereas one obtains a minimum of ${\cal D}(\omega)$ in this limit.

We are interested in the DC spin diffusion constant ${\cal D}_0={\cal D}(\omega \to 0) $ determined via Eq.~(\ref{dom}) by the offdiagonal matrix elements (ME) $j_{mn}$ at $|\epsilon_m-\epsilon_n| \to 0$. To connect to level sensitivity \cite{edwards72} we introduce finite flux into the exchange term via $J \to J \exp(\pm i \varphi)$ \cite{kohn64}. In the ergodic regime we can then verify the RMT relations that link diagonal and offdiagonal ME \cite{wilkinson90,castella96,dalessio16,schonle21}. Due to $\varphi \ne 0$, the time-reversal symmetry is broken and the relevant universality is that of the Gaussian unitary ensemble (GUE). The latter implies $Y = \overline{ | j_{mn}|^2}/\overline { j^2_{nn} } =1 $, provided the averaging is carried out over a narrow window of energies $\epsilon_n,\epsilon_m$ in the middle of the spectrum. This allows to evaluate the diffusion constant from either the offidiagonal or from the diagonal ME, where we take into account also statistical independence of ME $j_{mn}$ and energies $\epsilon_m-\epsilon_n$, being part of the eigenstate thermalization hypothesis (ETH) \cite{srednicki99,dalessio16}
\begin{eqnarray}
{\cal D}_0 &\simeq & \frac{\pi \overline{ | j_{mn}|^2} }{ L \tilde \chi_0} \frac{1}{N_{st}} \sum_{m\ne n} \delta(\epsilon_n - \epsilon_m)=\frac{ \pi\overline{ | j_{mn}|^2}}{L \tilde \chi_0 \Delta \epsilon} \label{odiag} \\
& \simeq & \frac{ \pi\overline { j^2_{nn} } }{L \tilde \chi_0 \Delta \epsilon} \label{diag} . 
\end{eqnarray}
Here, $ \Delta \epsilon= \overline{\Delta_n}$ is the average level spacing $\Delta_n = \epsilon_{n+1} - \epsilon_n$ or, equivalently, $(\Delta \epsilon)^{-1}$ is the MB density of states. Note that the diagonal ME are usually used to evaluate the ballistic (Drude) component of the conductivity in the integrable systems \cite{castella96,zotos97}, $D_D= \sum_n j^2_{nn}/(L N_{st}) = \overline{j^2_{nn} }/L$, while in ergodic (diffusive) systems $D_D$ decays exponentially with size. Equation~(\ref{diag}) demonstrates that in the finite systems obeying RMT, the diffusion constant and remnant ballistic component are related to each other via ${\cal D}_0= \pi D_D/(\tilde \chi_0 \Delta \epsilon)$.

Following arguments (originally introduced for noninteracting disordered systems) by Edwards and Thouless \cite{edwards72}, we investigate sensitivity $R$ of MB energies to changing of the boundary conditions from PBC to antiperiodic ones, or equivalently, to changing the flux by $ \delta \varphi=\pi/L$. $R$ can be expressed via diagonal ME as $j_{nn}= d\epsilon_n(\varphi)/d \varphi$ \cite{kohn64,castella96}, which combined with Eq.~(\ref{diag}) gives
\begin{eqnarray}
R \equiv \frac{ \delta \varphi \; \sqrt{\overline{(d\epsilon_n(\varphi)/d \varphi)^2}} }{\Delta \epsilon} 
= \frac{ \delta \varphi \; \sqrt{ \overline{ j^2_{nn} } } }{\Delta \epsilon} 
\simeq \sqrt{\frac{ \delta \varphi \; \tilde \chi_0 {\cal D}_0 }{\Delta \epsilon} }\,. \nonumber \\
\label{rrr}
\end{eqnarray}
Note that the quantities in Eq.(\ref{rrr}), i.e., level spacing, the matrix elements of the spin current as well as ${\cal D}_0$, should be evaluated within the same Hamiltonian, in particular for the same system size. This remark may be important for systems where ${\cal D}_0$ shows significant dependence on $L$. However, we do not observe such dependence in the present studies, at least not for the considered range of $W$.

It follows from Eq.~(\ref{rrr}) that in MB systems with finite DC diffusion, the level sensitivity parameter $R \propto 1/\sqrt{\Delta \epsilon} $ should grow exponentially with $L$. This relation can be considered as an alternative to the Thouless relation \cite{edwards72}, originally derived for noninteracting particles in random potentials. $R \gg 1$ implies that changing the boundary conditions induces multiple (avoided) level crossings, while $R \ll 1$ means effective insensitivity to boundary conditions, whereby $R_{th} \sim O(1)$ is a threshold value. Regime with $R < R_{th}$ can originate either from actual MBL or from the finite-size effects when DC diffusion is too small, i.e., ${\cal D}_0 < \Delta \epsilon$, according to Eq.~(\ref{rrr}). To differentiate between both scenarios it is crucial to follow the variation of $R$ with $L$. According to original formulation \cite{edwards72} of the Thouless criterion: the decreasing $R(L)$ implies localization, while its increase may be just a signature of finite-size effect, i.e., the system is too small. In any case, $R \sim O(1)$ gives also the bound on numerically accessible DC transport, $ {\cal D}_0 > {\cal D}_{\rm min} \sim \Delta \epsilon $. 

Most importantly, the Thouless criterion and Eq.~(\ref{rrr}) explain the $L$-shift of $W^{*}$, when the systems starts to exhibit deviations from RMT. This $L$-dependence emerges from Eq.~(\ref{rrr}) via the level spacing $\Delta \epsilon \propto 1/N_{st} \simeq \exp[-\log(2) L]$. Previous numerical studies of {\it random} spin chains \cite{barisic10,barisic16,steinigeweg16, prelovsek17,prelovsek21,herbrych22} clearly show the ${\cal D}_0 \propto \exp(-a W)$ dependence with constant $a \sim O(1)$. Using the exponential dependence of $\Delta \epsilon$ and ${\cal D}_0$, one finds that the Thouless criterion $R(W^{*})=R_{th}$ yields a linear drift $W^{*} \propto L$, well established in the numerical studies \cite{suntajs20,suntajs_bonca_20}. Below we demonstrate that the decay of ${\cal D}_0(W)$ in the QP chains is faster then exponential implying a sublinear drift $W^{*}(L)$, in agreement with previous the numerical studies \cite{aramthottil21}. We stress that linearity (or sublinearity) of $W^{*} (L)$ does not depend on a particular choice of threshold sensitivity $R_{th}$. 

\section{Numerical results for diffusion and ME relations  in finite systems}

In the following we verify numerically the above relations for the model in Eq. (\ref{hm}). The most convenient is the system with QP modulation (with $k=M/L$, $L$ and $M$ being relatively prime), where the MB density of states is featureless already for small $L\sim 20$. The analysis requires exact diagonalization (ED), where we study up to $L=18$ sites. For comparison, we consider also periodic potential $P=3$, where analysis should be performed with fixed total momentum which allows to reach $L=21$. In fact, periodic systems require more care since MB spectra reveal pronounced gaps at $W \gg 1$ (at reachable finite $L$). On the other hand, to evaluate ${\cal D}(\omega)$ and extract DC value ${\cal D}_0$ in considerably larger systems, we employ the microcanonical Lanczos method (MCLM) \cite{long03,prelovsek11,herbrych22} which allows to study systems with up to $L=28$ sites in $S^z_{tot}=0$ magnetization sector with chosen energies ${\cal E} \sim \overline H$.

\begin{figure}[tb]
\includegraphics[width=1.0\columnwidth]{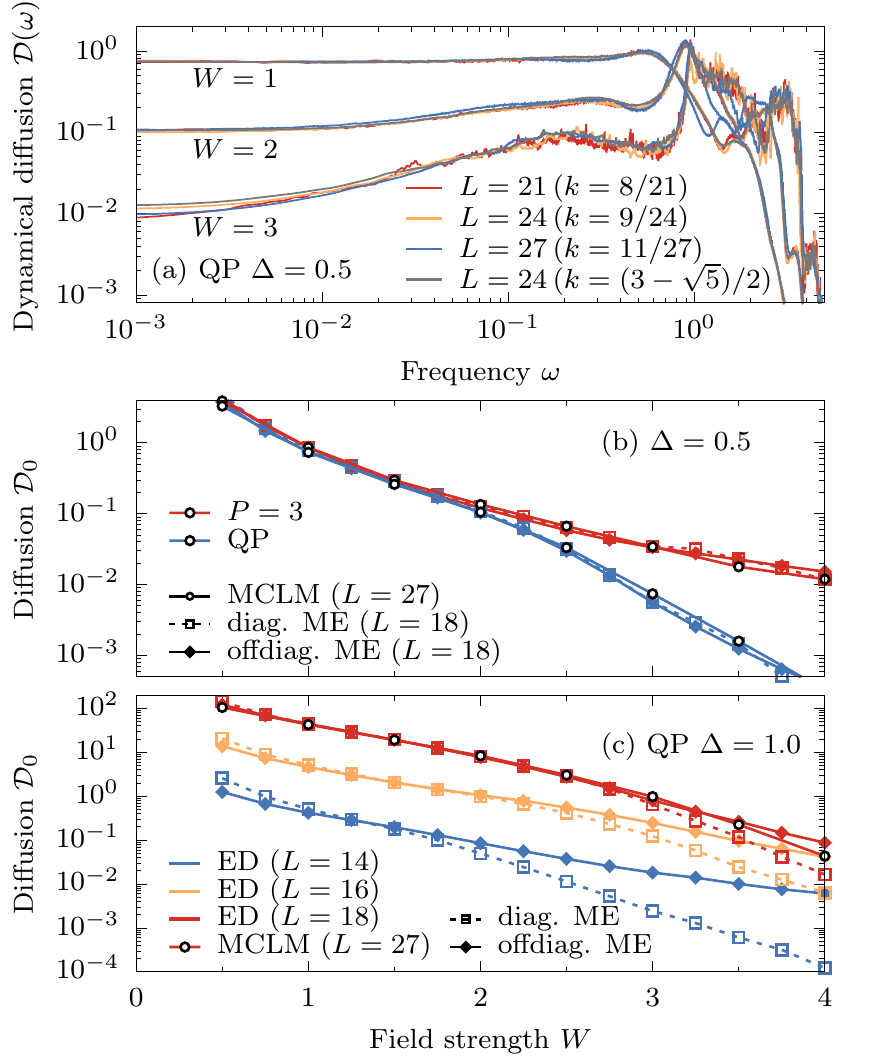}
\caption{(a) Dynamical diffusivity ${\cal D}(\omega)$ obtained via MCLM for QP chain with different system sizes $L$, compared also with the case when $k$ is an irrational number. (b-c) Spin diffusion constant ${\cal D}_0$ vs. potential strength obtained from offdiagonal ME [Eq.~(\ref{odiag})] and diagonal ME [Eq.~(\ref{diag})] via ED and compared to MCLM results for larger $L=27$ (periodic case with $P=3$) or $L=28$ (QP). (b) Results for $\Delta=0.5$ where ED is carried out for $L=18$ (QP) or $L=21$ (P=3). (c) QP chain with $\Delta=1$ and various $L$. For clarity, results for $L=16$ and $L=14$ are multiplied by factors $10$ and $10^2$, respectively.}
\label{fig1}
\end{figure}

We note that the dynamical diffusion in MBL-like systems has characteristic form\cite{karahalios09,barisic10,prelovsek17} ${\cal D}(\omega) \sim {\cal D}_0 + b|\omega|^{\alpha}$ with \mbox{$\alpha \gtrsim 1$}. Due to this unconventional $\omega$-dependence, the frequency-resolution of the applied method, $\delta \omega$, restricts the reachable DC values of diffusion constant to ${\cal D}_0 > {\cal D}_{min} \sim \delta \omega$. Therefore, it is crucial to have high frequency resolution which we achieve via large number of Lanczos steps $M_L$. The latter leads to $\delta \omega\sim \Delta E/M_L$, where $\Delta E$ is the whole MB energy span. We typically use $M_L \sim 4.10^4$ so that $\delta \omega \sim 4.10^{-4}$. Figure \ref{fig1} shows the MCLM results for $W \le 4$ where we find ${\cal D}_0 > \delta \omega$. We also note that the bulk diffusion constant should be formally obtained as $\lim_{\omega \to 0} \lim_{L \to \infty} {\cal D}(\omega)$ and that the these two limits may not commute in systems with anomalous transport, e.g., in integrable XXZ chain \cite{prelovsek2021}. In numerical approaches to MB quantum systems we are dealing generally with finite $L$ which also implies discrete, although very dense (exponentially for large $L$) spectra. In the case of normal diffusion the extrapolation of ${\cal D}_0$ does not depend on $L$ and extrapolation. Results in Fig.~\ref{fig1} are quite $L$-independent, allowing for a reliable estimation of ${\cal D}_0$ for the presented range of $W\le 4$. However, due to the above limitations, we don't formulate any claims concerning ${\cal D}_0$ for stronger disorders.

Results depicted in Fig.~\ref{fig1} indicate that the decay of ${\cal D}_0(W)$ in QP systems is even faster than exponential. It is best visible for weaker $\Delta=0.3, 0.5$ and $W>2$, where nonvanishing $\Delta$ is essential for stabilizing the diffusive transport \cite{aubry80}. Our numerical results for the diffusion constant can be well fitted by ${\cal D}_0 \propto W^{-\alpha W} = \exp(-\alpha W \ln W)$, however due to a limited range of accessible $W$, we can not benchmark this dependence against other possibilities. Our results for QP chains do not reveal any qualitative change of ${\cal D}_0(W)$ up to $W \simeq 4$, when we reach numerical limitations of MCLM. As a consequence, we do not see any clear indication for a transition to MBL at $W<4$. The studied range of $W$ covers the transition at $W^* \simeq 3$ reported in the literature \cite{naldesi16,barlev17,bera17, setiawan17,aramthottil21,sierant22} (obtained mostly from the gap ratio $\bar r$ in systems with $L \leq 22$).

Furthermore, our study reveals also that the regime of very small diffusion constant in QP systems can be approached via systems with simpler (integer) periodicities $P=2,3$, which evidently cannot exhibit MBL. In particular, numerical results reveal that the dependences on field strength $W$ start to deviate from that of the QP case only at large $W$, again putting restrictions on the MBL scenario. We discuss this issue at length in the Appendix~\ref{app1} for QP system and in the Appendix~\ref{app2} for the periodic case.

\section{Diffusion vs RMT} 
When deriving Eq. (\ref{rrr}) we have assumed that the diffusion constant can be obtained from Eq. (\ref{diag}), i.e., from RMT. Therefore, we first demonstrate that Eq.(\ref{diag}) indeed holds true even for quite substantial $W$. Fig.~\ref{fig1}(b) shows the diffusion constant for the QP system and a periodic chain ($P=3$) both with $\Delta=0.5$. It is evident that Eq.~(\ref{odiag}) involving offdiagonal ME, as well as the RMT-based Eq.~(\ref{diag}) with diagonal ME, accurately reproduce the MCLM results obtained for larger systems. The agreement holds for a broad range of ${\cal D}_0(W)$. For $P=3$ both relations, Eqs.~(\ref{odiag}) and (\ref{diag}), reproduce MCLM result up to largest considered $W \sim 4$, since ${\cal D}_0 > 10^{-2}$ and $R \gg1$ in the considered regime. On the other hand, it is expected that the deviations from RTM occur for finite QP chains with $W>W^{*}(L)$. Then, the GUE relation between diagonal and offdiagonal ME breaks down and Eq.~(\ref{diag}) systematically underestimates the results for the diffusion constant. We show such behavior in Fig.~\ref{fig1}(c) for $\Delta=1$. One observes also that $W^{*}$ increases with $L$. In the following we demonstrate, that the breakdown of RMT at $W^{*}$ and $L$-dependence of $W^{*}$ originate from the smallness of the diffusion constant relatively to the average level spacing.
 
It is instructive to follow besides $R$ and $Y$ also other indicators of RMT, i.e., the average gap ratio\cite{oganesyan07} $r = \overline r_n $ where $r_n =\mathrm{min} [\Delta_n,\Delta_{n+1}]/\mathrm{max}[\Delta_n,\Delta_{n+1}]$, with the GUE value\cite{atas13} $r \simeq 0.603 $, as well as the test of Gaussian distribution of ME, $Q= \overline{j_{nn}^4} / \left(\overline{j_{nn}^2}\right)^2$ with $Q=3$ for GUE \cite{wilkinson90,dalessio16}. In the following, we study the RMT indicators, in particular their size dependences for QP approximants $k=M/L \simeq 0.3$. Note that irrational $k$ are incompatible with the PBC considered in this work. Still, results in Fig. \ref{fig1}(a) show that the resulting ${\cal D}(\omega)$ is almost the same even when $k$ is taken as irrational (with some field inconsistency at the boundaries).

\begin{figure}[!htb]
\includegraphics[width=1.0\columnwidth]{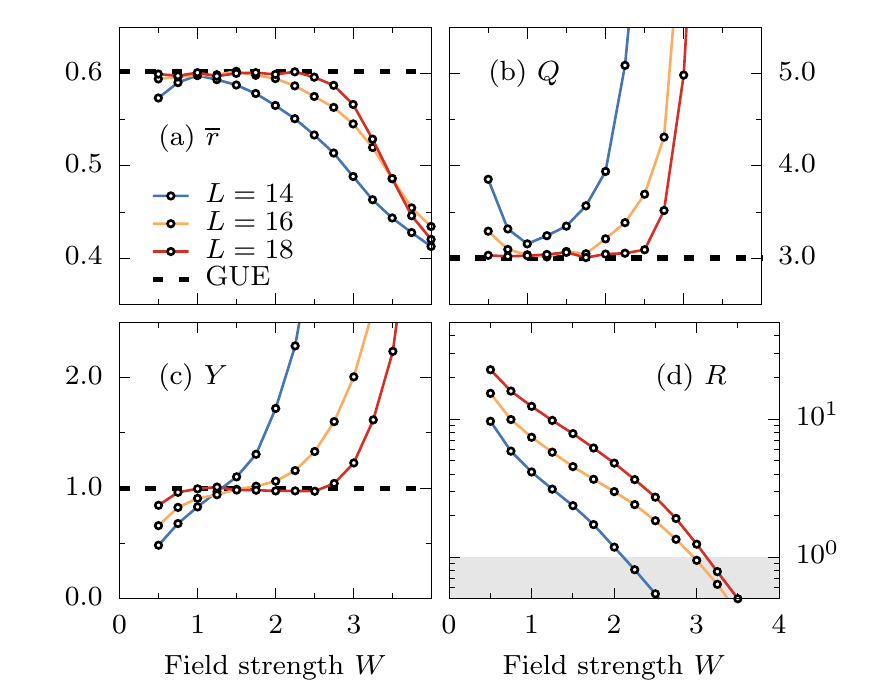}
\caption{ Level and current matrix-element criteria for the departure from the RMT universality vs. field strength $W$ for the QP case, calculated via ED for different system sizes $L =14-18$ and $\Delta=1$. (a) the gap ratio $\bar r$, (b) $Q$ from diagonal ME, (c) offdiagonal/diagonal ME ratio $Y$, and (d) level sensitivity parameter $R$. Dashed lines denote GUE values.}
\label{fig2}
\end{figure}

Figure~\ref{fig2} shows results for $(L=14, M= 5)$, $(L=16,M= 7)$ and $(L=18, M= 7)$. For fixed $L$, the gap ratio $\bar r$ as well as the indicators probing ME of the spin current, $Q$ and $Y$, start to deviate from the GUE predictions at the same $W \simeq W^{*}(L)$ at which the Thouless localization criterion $R(L) \sim O(1)$ \cite{edwards72}. On the other hand, $R$ still clearly increases with $L$, so apparent thresholds $W^*(L)$ does not represent MBL transitions, at least not at conjectured $W^* \simeq 3$ \cite{naldesi16,setiawan17,aramthottil21}.

Modifications of the energy levels induced by a local perturbation were previously discussed in Ref.~ [\onlinecite{serbyn15}]. In our approach we have followed the concept by Edwards and Thouless and studied the change of levels introduced via finite flux or equivalently via a phase modification of boundary conditions. It allowed us to derive a relation between the the flux sensitivity of energy levels, i.e., $d\epsilon_n(\varphi)/d \varphi $, and a transport response ${\cal D}_0$. In the case of a moderate disorders, conclusions from both approaches agree in that the level sensitivity parameter, $R$, as well as the corresponding quantity studied in Ref.~[\onlinecite{serbyn15}] increase with $L$. Results in Ref. ~[\onlinecite{serbyn15}] suggest that for stronger disorders the opposite happens. The latter regime is beyond the reach of the numerical methods applied in the present work since either ${\cal D}_0$ becomes smaller than the level spacing in system sizes accessible via ED or ${\cal D}_0$ becomes smaller than the energy resolution, $\delta \omega$, within the MCLM approach.

\section{Conclusions} 
Almost all up-to-date studies of MBL focused either on indicators derived from the level-statistics or on the transport properties. Here we establish the relation between both types of results. Namely, we derive the analogue of the Edwards-Thouless relation \cite{edwards72} that links the diffusion constant ${\cal D}_0$ to the structure of energy levels and the RMT-universality. While our numerical studies are carried out for QP systems, our results are generic and can be applied as well to random system. We show that the level-statistics, as well as indicators probing the matrix elements of spin current, yield similar thresholds potential (or disorder strength), $W^{*}$, for the breakdown of RMT in finite systems. The most informative measure is the level sensitivity parameter $R$, which marks the breakdown of RMT at $R (W^*) \sim \mathrm{O}(1)$. We show that the well established linear (for random systems) or sublinear (for QP chains) $L$-dependence of $W^{*}$ can be directly linked to the exponential-like decay of ${\cal D}_0(W)$. For the numerically accessible range of $W$, the level sensitivity in QP chains increases with $L$, not satifying the Thouless criterion for localization. Consequently, the crossover at $W^*$ in systems at present reachable via full ED are due to finite-size limitation (and to the smallness of ${\cal D}_0$). Clearly, this does not exclude a possibility of MBL transition at larger $W$, but at the same time puts limitations to its numerical detection. 

Our study also shows that the transport in QP chain, as directly relevant to the cold-atom experiments \cite{schreiber15,luschen17}, can be well approached via transport properties obtained for a sequence of models with simple periodic fields partially resembling the case of a noninteracting QP system \cite{szabo2018}. Still, we find that even in a simple periodic system, e.g., with periodicity $P=3$, diffusion ${\cal D}_0(W)$ follows (even quantitatively) its dependence as in QP case up to substantial $W$. Only at large $W$ we observe qualitatively different behavior of QP and periodic systems, again having the consequences for the scenario of the potential MBL transition. 

\noindent{\it Acknowledgments.}
M.M. acknowledges the support by the National Science Centre, Poland via projects 2020/37/B/ST3/00020. P.P. acknowledges the support by the project N1-0088 of the Slovenian Research Agency. The numerical calculation were partly carried out at the facilities of the Wroclaw Centre for Networking and Supercomputing.

\appendix

\section{Diffusion constant}
\label{app1}

\subsection{Phase $\phi_0$ dependence}
\begin{figure}[!htb]
\includegraphics[width=1.0\columnwidth]{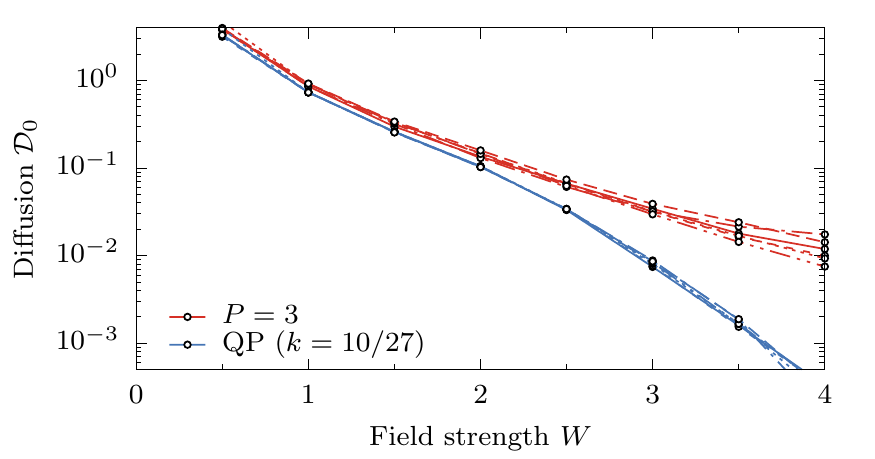}
\caption{DC diffusion ${\cal D}_0$ vs. field strength $W$, calculated with MCLM for $L=27$ at $\Delta =0.5$ for a periodic $P=3$, and QP chains. The presented data depict five different choices of  phase shift $\phi_0$.}
\label{figS3}
\end{figure}

Let us first comment on the dependence of the ${\cal D}_0$ results on the phase shift $\phi_0$ in local fields $h_i=(W/2) \cos( 2\pi ki + \phi_0)$. We note that in majority of previous studies on the QP case, results (predominantly on $\bar r$) have been averaged over $\phi_0$ \cite{naldesi16,barlev17,bera17, setiawan17,aramthottil21,sierant22} to improve statistics in finite systems. In Fig.~\ref{figS3} we present MCLM results for ${\cal D}_0$ at $\Delta=0.5$ and $L=27$ for the case of QP field ($k=10/27$) as well as for $P=3$ with various nonequivalent phases $\phi_0$. It is evident that results match well even quantitatively up to largest $W \sim 4$. This is expected for QP case, since ${\cal D}_0$ is not affected even by changing between different QP approximants $k=10/27, k=11/27$. On the other hand, it is evident from the data presented in Fig.~\ref{figS3} that the phase $\phi_0$ can influence results for systems with integer period $P$. While for $P=2$ only $\phi_0=0$ is meaningful, the variation for $P=3$ is somewhat larger than for QP case, but nevertheless quite weak and does not change any qualitative conclusions.

\subsection{$\Delta$ dependence}

\begin{figure}[!htb]
\includegraphics[width=1.0\columnwidth]{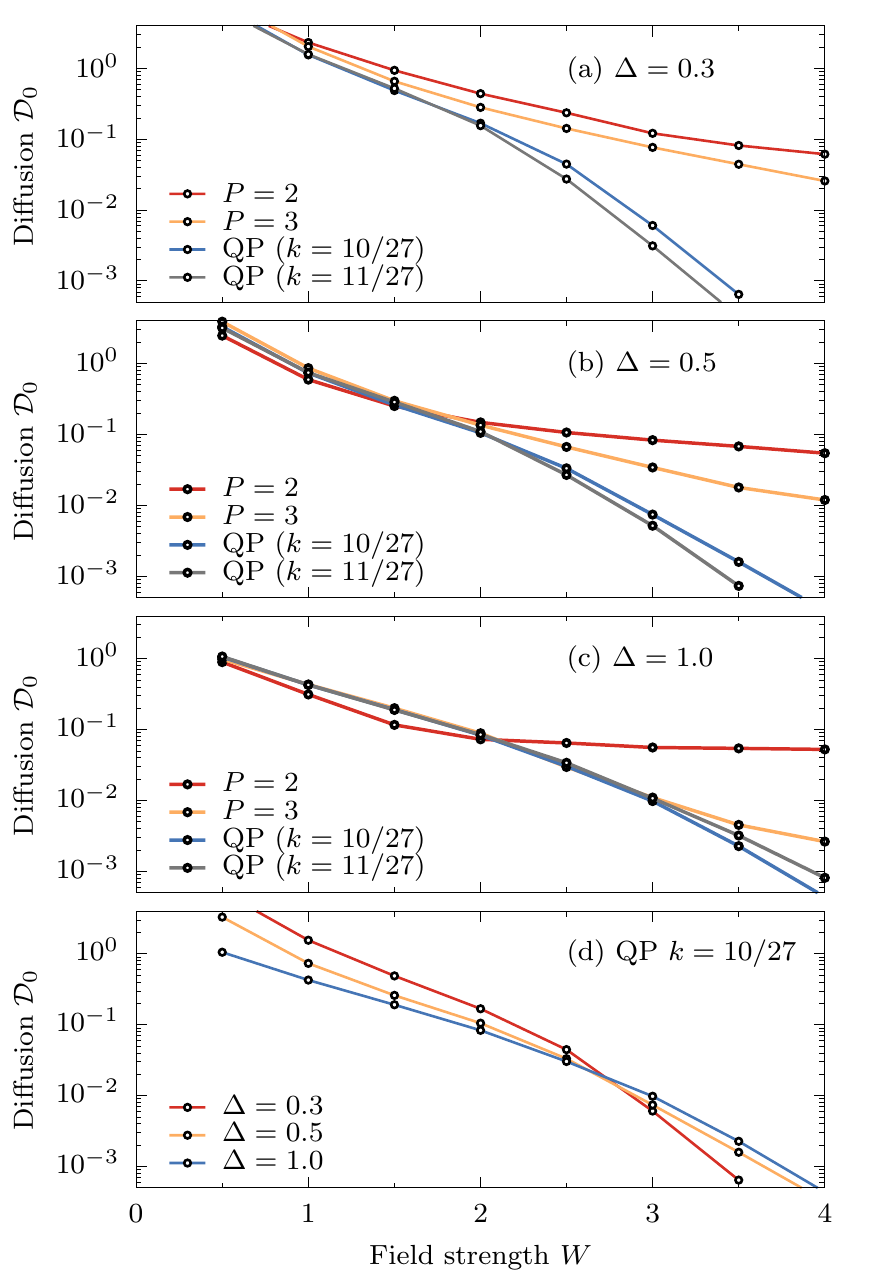}
\caption{DC diffusion ${\cal D}_0$ vs. field strength $W$ as calculated with MCLM for $P=2$, $P=3$ and two QP approximants, and (a) $\Delta=0.3$, (b) $\Delta=0.5$, (c) and $\Delta=1.0$. Panel (d) depicts QP ($k=10/27$) case and different anisotropies $\Delta=0.3, 0.5, 1.0$.}
\label{figS4}
\end{figure}

We now turn to the results for DC diffusion ${\cal D}_0$ obtained via MCLM for largest reachable $L=27$ and QP. In Fig.~\ref{figS4} we present results for ${\cal D}_0$ vs. potential strength $0.5 \leq W \leq 4$ for few anisotropies, i.e., the most studied isotropic case $\Delta=1$, and more modest $\Delta=0.5$ and $0.3$. In the same figure, we present also results for periodic potential $P =2,3$. In addition, we compare in Fig.~\ref{figS4}(d) MCLM results for different $\Delta=0.3, 0.5, 1$ for the QP case.

We first note a large span of results for the diffusion constant, $ 10^{-3} \lesssim {\cal D}_0 \lesssim 10^0$, in particular for the QP case. The variation ${\cal D}_0(W)$ can be separated into different regimes.

\noindent (i) At weak $W \lesssim 1$ we are dealing with perturbed integrable system, where $W \ne 0$ introduces scattering, so that ${\cal D}_0 \propto 1/W^2$ (as well as ${\cal D}_0 \propto 1/\Delta^2$) is expected independently of periodicity of $h_i$.

\noindent (ii) For stronger potentials, the results for periodic chains deviate from the QP case, with the differences more pronounced for smaller $P$. Moreover, the larger $P$ or $\Delta$ are, the larger is $W$ when these deviations become significant. Strong $\Delta$-dependence of ${\cal D}_0$ in QP case is expected, since such system at $\Delta =0$ is localized for $W > 2$ \cite{aubry80}. On the other hand, for $P=2$ the diffusion constant ${\cal D}_0$ is rather independent of $\Delta$, at least for modest $\Delta >0$ considered here. For $P=3$ large-$W$ regime is less straightforward, since increasing $\Delta$ reduces ${\cal D}_0$, but with dependence on $W$ being of power-law type.

\noindent (iii) Eventually for large $W$, the diffusion constant in the periodic systems follows a power-law dependence ${\cal D}_0 \propto W^{-\zeta}$. Clearly, such asymptotic power-law decay of ${\cal D}_0( W)$ is consistent with the expectation that there is no MBL in periodic systems, hence ${\cal D}_0$ may be small but nonzero.

\noindent (iv) For QP fields the dependence on $\Delta$ is inverted for $W\gtrsim 2$, as evident in Fig.~\ref{figS4}(d). This is consistent with the fact, that at $\Delta=0$ the model is equivalent to the chain of noninteracting fermions in QP potential with localized states for $W> 2$ \cite{aubry80}. Only finite $\Delta >0$ can then induce ${\cal D}_0 >0 $, with a nontrivial dependence on $\Delta$.

\section{Periodic system}
\label{app2}

\subsection{Effective two-band model }

Let us try to explain the diffusion in simple $P=2$ system. At $\Delta=0$ the model, Eq.~\ref{hm}, maps on the chain of noninteracting fermions corresponding to two bands with dispersion $E^\pm_k =\pm \sqrt{ J^2 \cos^2{k} + W^2/4}$. Since we are interested in the behavior at large $W>W_0 = 2$, at $\Delta \leq 1$ the field term with $W$ represents the largest scale in the problem, and we have two well separated (narrow) bands, $ E^\pm_k \sim \pm W/2 \pm J^2 \cos^2k/W$. 

\begin{figure}[!b]
\includegraphics[width=1.0\columnwidth]{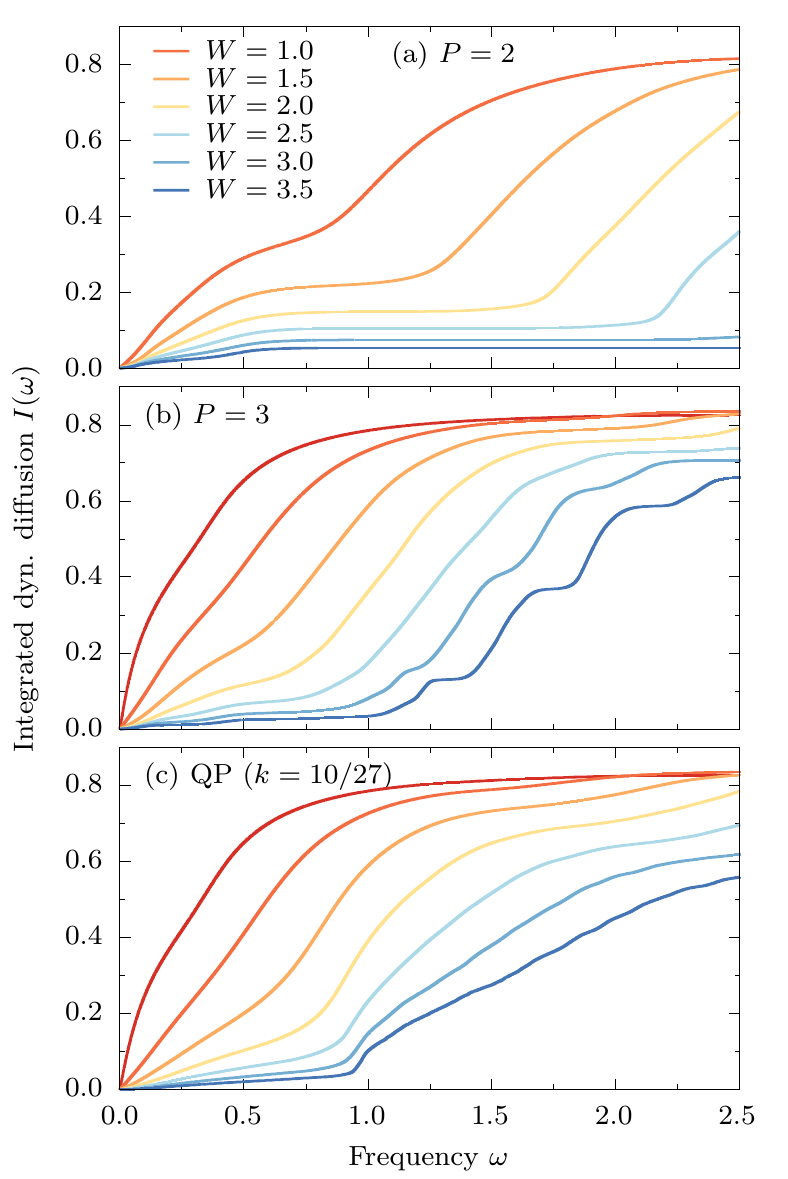}
\caption{Integrated dynamical-diffusion spectra $I(\omega)$ calculated with MCLM for the spin chain with $\Delta=0.5$ and various $W$ for (a) periodicity $P=2$ ($L=28$) (b) periodicity $P=3$ ($L=27$), and (c) QP approximant $k=10/27$ with $L=27$.}
\label{figS6}
\end{figure}

\begin{figure}[!b]
\includegraphics[width=1.0\columnwidth]{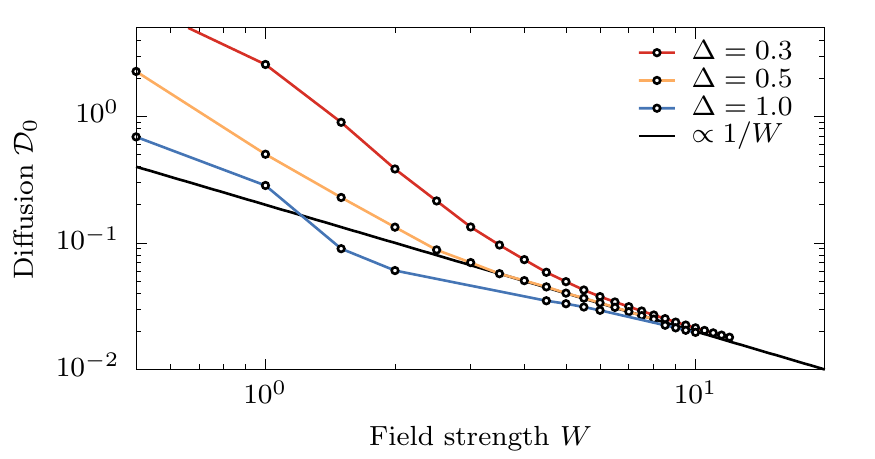}
\caption{Diffusion constant ${\cal D}_0$ for $P=2$ with different anisotropies $\Delta=0.3, 0.5, 1$, evaluated with ED on $L=16$ sites. At large $W \gg 1$ results confirm asymptotic scaling ${\cal D}_0 \propto 1/W$. }
\label{figS7}
\end{figure}

Following only the low-energy processes, at $\Delta >0$ we deal with an effective ladder model, with two sorts of (spinless) fermions $c_{lj}$ with $j = [1, L/2]$ and $l=a,b$, i.e., electrons hopping only on the same rail (within the same band) via a second-order process:
\begin{equation}
\tilde H = t_2 \sum_{j,l=a,b} (c^\dagger_{l,j+1} c_{lj} + \mathrm{H.c}) + 
J \Delta \sum_j n^a_j ( n^b_j + n^b_{j+1} ), \label{heff}
\end{equation}
with $t_2 = J^2/(4W)$. Note that the effective model conserves the number of particles within each band/rail $N^l = \sum_j n^l_j$. The transport properties of Hamiltonian~(\ref{heff}) are still nontrivial since the emerging effective interaction $\tilde \Delta =J \Delta /t_2= 4 \Delta W/J$ can become large for $W > W_0$, i.e., $\tilde \Delta >1$ even for modest $\Delta \leq 1$. The validity of the effective model in Eq.~(\ref{heff}) can be checked by the low-$\omega<1$ sum rule for integrated $I(\omega)=\int_0^\omega \mathrm{d}\omega'\,{\cal D}(\omega') $. We note that the whole sum rule is $I_0 = I(\infty) = \pi J^2/4$, while the reduced one can be extracted from the renormalized hopping term in Eq.~(\ref{heff}),
\begin{equation}
\frac{I_2}{I_0} \propto \frac{16 t_2^2}{J^2} = \frac{4J^2}{W^2} \ll1, \qquad \mathrm{for}~~ W \gg W_0. \label{i2}
\end{equation}
It is evident from the results presented in Fig.~\ref{figS6}(a) ($P=2$ case) that for $W>1$ the spectra reveal pronounced plateaus in $I(\omega)$, consistent with Eq.~(\ref{i2}) and corresponding to the band gaps in dynamical ${\cal D}(\omega)$. Note however that the effective model does not offer a simple explanation for the DC diffusion at large $W$. Due to its resemblance to the Hubbard model and its transport properties at high $T$ \cite{perepelitsky16,kokalj17,ulaga21}, we are dealing with incoherent diffusion within each band, i.e., with minimal mean free path $l_0 \sim 2$ and the effective velocity $v_\mathrm{eff} \propto t_2$. As a consequence, ${\cal D}_0 \propto v_\mathrm{eff}\, t_2 \propto 1/W$, which is consistent with numerical findings presented in Fig.~\ref{figS7}.

Fig.~\ref{figS6}(b) and Fig.~\ref{figS6}(c) show $I(\omega)$ for $P=3$ and QP cases, respectively. Here we use $\Delta=0.5$, where the appearance of gaps is more pronounced than for $\Delta=1$. For periodicity $P=3$ the gap starts to emerge for $W \geq 1.5$, the value which is larger than for smaller periodicity $P=2$. Also, the gap structure is less pronounced since at the same $W$ gaps are smaller. Consistent with consideration of the related effective model for $P=3$, the low-$\omega$ sum rule for $I(\omega)$ has stronger dependence on $W$. Although it is evident that the structure of ${\cal D}(\omega<1)$ is highly nontrivial, the results for DC diffusion constant at large $W \gg 1$ are (in analogy with $P=2$) consistent with the power-law dependence ${\cal D}_0 \propto 1/W^\zeta$. In contrast to periodic systems, the QP case as shown in Fig.~\ref{figS6}(c), does not reveal pronounced gaps at any $W$, but a rather uniform variation for $\omega <1$. Evident is also a resonance at $\omega \sim 1$ for which we do not have a simple explanation.

\begin{figure}[!t]
\includegraphics[width=1.0\columnwidth]{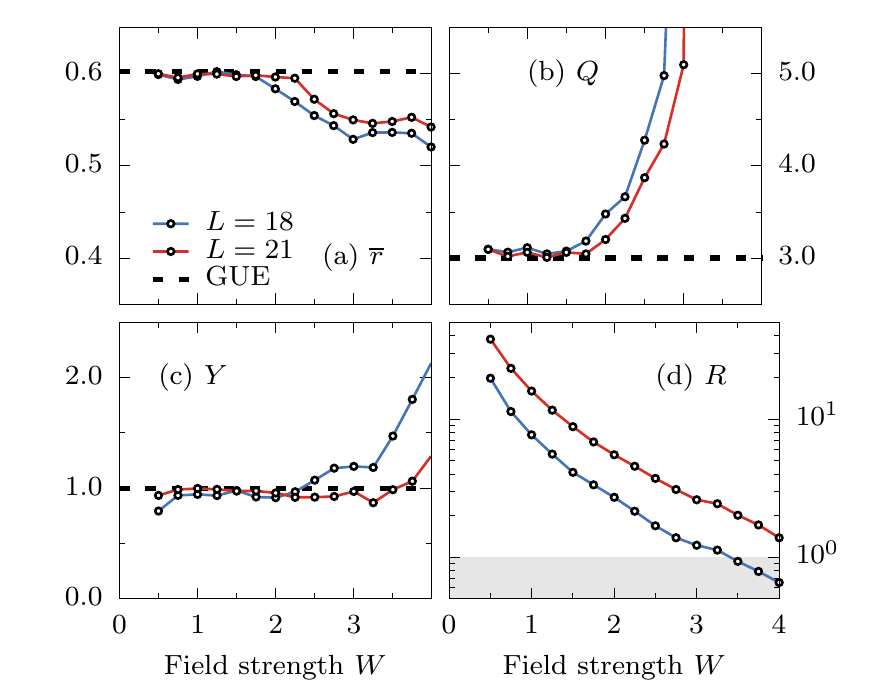}
\caption{Level and current matrix-element criteria for the departure from the GUE universality vs. field strength $W$ for the periodic field $P=3$ case, calculated via ED for $L=18,21$. Panel (a) depicts the gap ratio $\bar r$, (b) $Q$ from the diagonal ME , (c) offdiagonal/diagonal ratio $Y$, and (d) level sensitivity parameter $R$. Dashed curves denote the GUE values.}
\label{figS1}
\end{figure}

\subsection{Level and current matrix-element statistics}

\begin{figure}[!b]
\includegraphics[width=1.0\columnwidth]{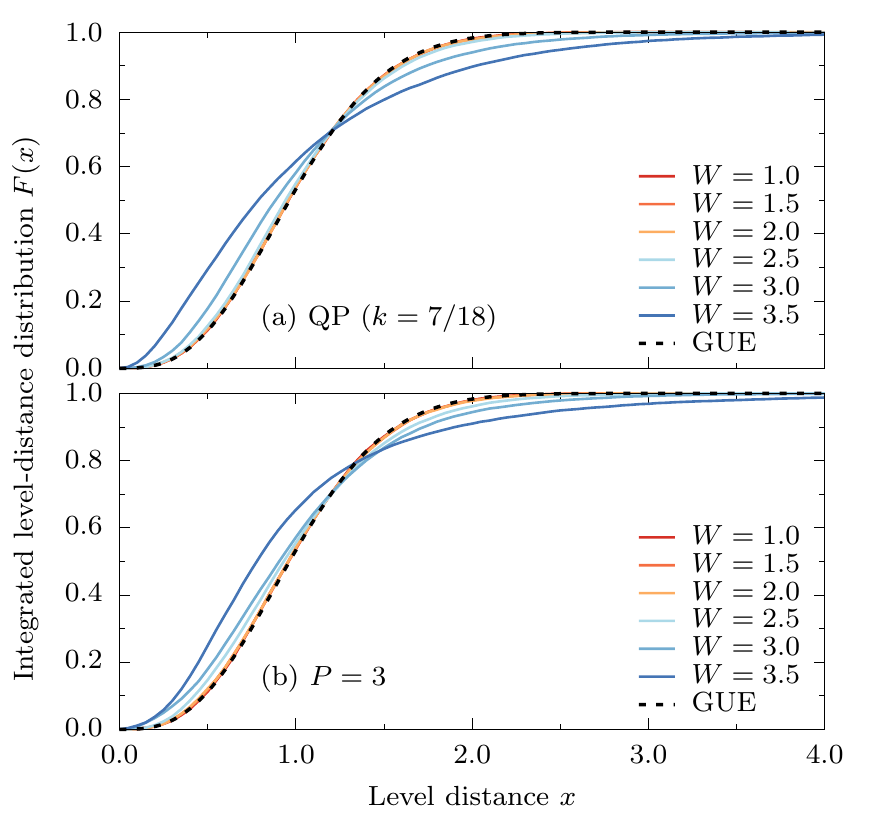}
\caption{Integrated level-distance distribution $F(x)$ for different $W$, as calculated via ED for: (a) the QP system with $\Delta=1$ on $L=18$ sites, and (b) for $P=3$ system with $\Delta=0.5$ on $L=21$ sites. Dashed line denotes the GUE dependence from Eq.~(\ref{gue}).}
\label{figS2}
\end{figure}

In analogy to the QP case, as presented in the main text, we analyse in the following also the RMT indicators for the periodic case $P=3$. In order to eliminate additional symmetries due to finite periodicity, the analysis has been performed in sectors with fixed translation wavevector. We present in Fig.~\ref{figS1} results for the gap ratio $\bar r$, ME criteria $Q$ and $Y$ as well the level sensitivity parameter $R$, as function of $W$ for fixed $\Delta=0.5$ and two sizes $L=18,21$. Since $P=3$ system should remain ergodic even for large $W$ (in $L \to \infty$ limit) and due to large ${\cal D}_0$, one would expect the GUE universality in a broader regime of $W$ than in QP (for given $L$). Indeed, $\bar r$ and $Y$ do not deviate strongly from GUE values even for largest $W \sim 4$. This is consistent with $R >1$ remaining beyond the threshold even for $W \sim 4$, at least for $L=21$. On the other hand, $Q$ starts to deviate from Gaussian value already for intermediate $W \sim 2$. This can reconciled with the observation that for $W\sim 2$ the spectrum of the model start to reveal the separation into bands, having for modest $L$ the consequence of finite-size gaps in MB spectra (see the discussion later on).

Another standard measure for the validity of RMT is the normalized level-distance distribution ${\cal P}(x)$, where $x=(\epsilon_{n+1}-\epsilon_n)/\Delta \epsilon$. Since we consider here the complex Hamiltonian, Eq.~(\ref{hm}) of the main text with added flux, one expects within the GUE the ${\cal P}(x)$ of the form \cite{dalessio16},
\begin{equation}
{\cal P}(x) = \frac{32x^2}{\pi^2} \exp(- 4 x^2/\pi). \label{gue}
\end{equation}
In Fig.~\ref{figS2} we present the result for integrated distribution function \mbox{$F(x) =\int^x_0\mathrm{d}x'\, {\cal P}(x')$} for the QP case with $\Delta=1$ (considered also in Fig.~\ref{fig2} in the main text) and $P=3$ for $\Delta =0.5$ (corresponding to Fig.~\ref{figS1}). The results are calculated via ED method for the $L=18$ and $L=21$, respectively. Conclusions following from these results are quite consistent with other RMT criteria: (i) for QP system deviations from Eq.~(\ref{gue}) become visible at $W>3$, where also other criteria deviate from GUE (for the same $L$). (ii) On the other hand, $F(x)$ for $P=3$ starts to deviate from GUE in spite of $R > 1$ (in analogy to $Q$ in Fig.~\ref{figS1}). The latter can be explained by quite pronounced gaps in MB eigenvalues and would require more careful unfolding of MB spectra. 

\bibliography{manuqp}

\end{document}